\documentclass[12pt]{article}
\usepackage{graphicx}
\usepackage{amssymb}
\usepackage{amsmath}
\usepackage{xcolor}
\usepackage{soul}
\setstcolor{red}
\usepackage{cite}

\usepackage{array}
\usepackage{subcaption}
\usepackage{booktabs}

\makeatletter

\@addtoreset{equation}{section}
\makeatletter

\DeclareMathOperator{\sgn}{sgn}

\newcommand{\bc}{\textcolor{blue}}

\begin{document}

\begin{titlepage}
\vfill
\begin{flushright}
\end{flushright}

\vfill
\begin{center}
\baselineskip=16pt
{\Large\bf 
Geodesic Structure of Naked Singularities \\ in AdS$_3$ Spacetime\\
}
\vskip 0.5cm
\vskip 10.mm
{\bf Cristi{\'a}n Mart\'{\i}nez${}^{a}$,  Nicol\'as Parra${}^{b}$, Nicol\'as Vald\'es${}^{b}$ and Jorge Zanelli${}^{a}$} \\
\vskip 1cm
{
    ${}^a$ Centro de Estudios Cient\'{\i}ficos (CECs), Av. Arturo Prat 514, Valdivia, Chile. \\
     ${}^b$ Departamento de F\'{\i}sica, FCFM, Universidad de Chile,  Blanco Encalada 2008,\\ Santiago, Chile. \\
     \vskip 0.2cm
	\texttt{\footnotesize{martinez@cecs.cl, nicopave@gmail.com, n.valdes.meller@gmail.com, z@cecs.cl}}
} 
\vspace{6pt}
\end{center}
\vskip 0.2in
\par
\begin{center}
{\bf Abstract}
\end{center}
\begin{quote}
We present a complete study of the geodesics around naked singularities in AdS$_3$, the three-dimensional anti-de Sitter spacetime. These stationary spacetimes, characterized by two conserved charges --mass and angular momentum--, are obtained through identifications along  spacelike Killing vectors with a fixed point. They are  interpreted as massive spinning point particles, and can be viewed as three-dimensional analogues of cosmic strings in four spacetime dimensions. The geodesic equations are completely integrated and the solutions are expressed in terms of elementary functions. We classify different geodesics in terms of their radial bounds, which depend on the constants of motion. Null and spacelike geodesics approach the naked singularity from infinity and  either fall into the singularity or wind around and go back to infinity, depending on the values of these constants,  except for the extremal and massless cases for which a null geodesic could have a circular orbit. Timelike geodesics never escape to infinity and do not always fall into the singularity, namely, they can be permanently bounded between two radii. The spatial projections of the geodesics (orbits) exhibit self-intersections, whose number is particularly simple for null geodesics. As a particular application, we also compute the lengths of fixed-time spacelike geodesics of the static naked singularity using two different regularizations.

\vfill
\vskip 2.mm
\end{quote}
\end{titlepage}

\section{Introduction} 
Shortly after the BTZ black hole in three-dimensional spacetime was discovered \cite{BTZ1,BTZ2}, its geodesic structure was studied in detail \cite{Farina:1993xw,Cruz1994}. The study of those geodesics not only gave insight into the geometrical properties of the BTZ spacetime, but also led to some interesting applications \cite{strings, interpretEL, flows, collision}.

The same BTZ metric, with mass $M$ and angular momentum $J$, when continued to negative values of $M$ describes other interesting geometries. In the static case $J=0$, for $0>M>-1$ the resulting three-dimensional spacetime is a conical geometry with deficit angle and a naked singularity at the apex of the cone \cite{MZ}. Naked singularities (NS) correspond to point particles, while geometries with angular excesses may be interpreted as antiparticles  \cite{DJT,DJ}. These naked singularities can be viewed as the $2+1$ dimensional analogues of cosmic strings in $3+1$ dimensions and have been an object of extensive study in the past \cite{deserjackiw, sousajackiw }. Here we study the geodesics around these BTZ naked singularities (NS), extending to $M<0$ what is already known about the black holes. 

From a geometric perspective, the NS in three-dimensional AdS spacetime can be obtained by identifying points along a rotational Killing vector. More precisely, one identifies with rotations on two independent planes in $\mathbb{R}^{(2,2)}$. Identifications in AdS$_3$ have been used to describe black hole formation \cite{BHcreate} and a construction of time machines \cite{dedeo}. In this paper we present all possible geodesics around of BTZ cones. Our analysis may contribute to an understanding of some aspects of current interests: specifically,  these geodesics could be useful for discussing entanglement entropy \cite{RT, lengthsRT}, and for recent studies of quantum backreaction on naked singularities \cite{ dress1, Casals:2016odj,  dress2, CFMZ}. 

This paper is organized as follows. In Section 2 we review the NSs and discuss how to obtain them by identifications on the covering AdS$_3$ space  embedded as a pseudo-sphere in $\mathbb{R}^{2,2}$. In Section 3 we use conserved quantities along geodesics to find the first order geodesic equations in NS spacetimes, and then through re-scaling write the equations in a convenient form. In Section 4,  we present  solutions to the radial equations and corresponding bounds for the different types of geodesics. We note that all null geodesics escape to infinity or fall into the singularity, except for a  special case which allows circular orbits. Similarly, spacelike geodesics either have both ends at infinity, or one end at infinity and the other at the singularity. Meanwhile, all timelike geodesics are bounded: they either orbit the NS at finite radius or fall into it. Section 5 deals with the spatial projections of the geodesics. We find exact analytic solutions, plot representative orbits and discuss their qualitative behavior.  In Section 6, we analyze an interesting property of the geodesics arising from the results of Section 5: null, spacelike, and timelike  geodesics can all intersect themselves, and we calculate the number of self-intersections. Section 7 presents two more properties: the time behavior of geodesics, and the lengths of spacelike geodesics. The last section contains a summary and discussion of the main results. 

\section{The NS spacetime}  
Although all vacuum solutions of the three-dimensional Einstein equations with negative cosmological constant $\Lambda$ are constant curvature spacetimes locally isometric to AdS$_3$, there exist geometries globally distinct from AdS$_3$, including black holes. This is the case of the family of BTZ geometries described by the stationary line element
\begin{equation} \label{BTZmetric}
ds^2= -\left(\frac{r^2}{\ell^2}-M\right)dt^2 -Jdt d\theta + \left(\frac{r^2}{\ell^2}-M+\frac{J^2}{4r^2}\right)^{-1}dr^2+r^2 d\theta^2,
\end{equation}
where $\ell^2=-\Lambda^{-1}$, $-\infty<t<\infty$, $0<r<\infty$, and $0\leq \theta\leq 2\pi$. Here the mass $M$ and angular momentum $J$ are integration constants\footnote{We set the three-dimensional Newton constant as $G=1/8$.}. Depending on the values of $M$ and $J$ various spacetimes emerge from the BTZ metric \eqref{BTZmetric}, which are summarized in Table 1.
\begin{table}[ht]
\centering
\caption{BTZ geometries for different $M$ and $J$}
\begin{tabular}{c|c}
\hline 
\bf{ $\boldsymbol{M}$ - $\boldsymbol{J}$ regions}  &  \bf{Geometries} \\
\hline \hline
$M > 0$ and $|J| < M \ell$ & Black holes \\ \hline 
$M > 0$ and  $|J| = M \ell$ &  Extremal black holes \\ \hline
$M < 0$ and $|J| < -M \ell$ & Naked singularities\\ \hline 
$M < 0$ and $|J| = -M \ell$ & Extremal naked singularities\\ \hline 
$M = 0$ and $J=0 $ & Massless BTZ geometry\\ \hline
$M = -1$ and $J=0 $ & AdS$_3$ vacuum \\
\hline 
\end{tabular}
\end{table}

Here we are interested in the naked singularities (with $|J|\leq |M|\ell$), namely, the geometries without an event horizon, which correspond to spacetimes with nonpositive mass $M \le0$. The only exception is the case $M=-1, J=0$, which is the AdS$_3$ spacetime. NSs can be obtained by identifications on the universal covering space CAdS$_3$ \cite{MZ}. We present below a brief review of this construction.

Consider CAdS$_3$ as the set of points $X^a=\left(X^{0},X^{1},X^{2},X^3\right)$ of the pseudo-sphere embedded in $\mathbb{R}^{2,2}$ defined by
\begin{equation} \label{pseudosphere}
-(X^0)^2+(X^1)^2+(X^2)^2- (X^3)^2=-\ell^2. 
\end{equation}
This embedding can be parametrized with coordinates $(t, r, \theta)$ on the hypersurface defined by  \eqref{pseudosphere}, which yields the induced metric \eqref{BTZmetric}. The Killing vector $\Theta= \partial_{\theta}$ is chosen as the identification vector and is written as a linear combination of the $so(2,2)$ generators $J_{a b}:=X_b \partial_a-X_a \partial_b$,\footnote{Here $X_a=\eta_{ab}X^b$, with $\eta_{ab}=\text{diag}(-,+,+,-)$.}
\begin{equation}
\Theta= \frac{1}{2} \omega^{ a b} J_{a b}  =\frac{\partial X^a}{\partial \theta}  \partial_a =\partial_{\theta},
\end{equation}
where the antisymmetric matrix $\omega^{ a b}$ characterizes the identification in terms of the $so(2,2)$ generators. Then, the action of the matrix $H = e^{2\pi \Theta}$ on the coordinates of the embedding space is 
\begin{equation}
H^{a}\,_{b} \, X^b(t, r, \theta) = X^a(t, r,\theta +2\pi). 
\end{equation}
The explicit form of the embeddings for the different geometries and the corresponding identification matrices $H$ can be found in Refs. \cite{MZ,CFMZ}. The different identification vectors $\Theta$ are shown in Table \ref{tablevectors}, where we have defined
\begin{equation}
b_{\pm} = \frac{1}{2}\left(\sqrt{-M+J/\ell}\pm\sqrt{-M-J/\ell}\right).\label{bpm}
\end{equation}
\begin{table}[ht!] 
\centering
\caption{Identification Killing vectors $\Theta$ in terms of $so(2,2)$ generators for different NS spacetimes }
\begin{tabular}{c|c}
\hline 
\bf{Killing vector $\boldsymbol{\Theta}$} & \bf{Geometry} \\ \hline \hline 
 $b_+ J_{2 1}+b_- J_{30}$ & Generic NS ($0<M \ell<-|J|$)\\ \hline
$\sqrt{-M/2} (J_{03}-J_{12})-\frac{1}{2} (J_{01}+J_{03}+J_{12}-J_{23})$ & Extremal NS ($M \ell=-|J|)$ \\ \hline 
$J_{12}-J_{13}$  & Massless BTZ geometry ($M=J=0$) \\ \hline
\end{tabular} \label{tablevectors}
\end{table}

As Table 2 displays, the non-extremal NS is obtained by an identification by a Killing vector formed by two rotations. Note that for the extremal and massless cases, the Killing vectors contain rotations and boosts that are not limiting cases of the generic form.

\section{Geodesic equations} 

The NS spacetimes have two Killing vectors $\xi=\partial_t$ and $\Theta=\partial_\theta$. They provide two conserved quantities along the geodesic motion, $E=-\xi_{\mu}\dot{x}^{\mu}$ and $L=\Theta_{\mu}\dot{x}^{\mu}$, respectively, where $\dot{x}^{\mu}=dx^{\mu}/d\lambda$ is tangent to the geodesic with affine parameter $\lambda$. This allows us to obtain the first integrals
\begin{align}
\label{tdot}
\dot{t}&= \frac{E r^2 - JL/2}{r^2\left(\frac{r^2}{\ell^2}-M+\frac{J^2}{4r^2}\right)},\\
\label{thetadot}
\dot{\theta} &= \frac{(r^2/\ell^2 - M)L +J E/2}{r^2\left(\frac{r^2}{\ell^2}-M+\frac{J^2}{4r^2}\right)}.
\end{align}
 Since the velocity can be normalized as $\dot{x}^{\mu}\dot{x}_{\mu}=-\varepsilon$ (with $\varepsilon=0$ for null geodesics, $\varepsilon>0$ for timelike geodesics, and $\varepsilon<0$ for spacelike geodesics), one gets
\begin{equation}
\label{rdotgen}
 r^2\dot{r}^2 = -\varepsilon r^2\left(\frac{r^2}{\ell^2}-M+\frac{J^2}{4r^2}\right)+\left(E^2-\frac{L^2}{\ell^2}\right)r^2 +L^2 M - JEL.
\end{equation}

Equations (\ref{tdot}-\ref{rdotgen}) are exactly the same for black holes ($M>0$) \cite{Cruz1994} and naked singularities ($M<0$). However, the denominators of $\dot{\theta}$ and $\dot{t}$ vanish at the BH horizons, while for NS they are positive definite. Consequently, the geodesics around a NS are drastically different from those in the BH case.

Using (\ref{bpm}) it is convenient to write 
\begin{equation}
r^2\left(\frac{r^2}{\ell^2}-M+\frac{J^2}{4r^2}\right)=\ell^2 \left(\frac{r^2}{\ell^2}+b_{+}^2\right) \left(\frac{r^2}{\ell^2}+b_{-}^2\right), 
\end{equation}
and since $M<0$ and $\ell \neq 0$, it is useful to introduce the following quantities\footnote{The massless BTZ spacetime is considered separately in \ref{Massless BTZ Geodesics}.}
\begin{align}
u=r^2/(-M \ell^2), \quad a= J/ (-M\ell).
\end{align}
Furthermore, for $M\neq 0$ we use the rescaled quantities
\begin{align}
    \tilde{L}=L/&(-M\ell), \quad \tilde{E}=E/(-M),  \quad \tilde{\varepsilon}=\varepsilon/(-M), 
\quad \tilde{\lambda}=\lambda/\ell, \nonumber \\
&\tilde{b}_{\pm}=b_{\pm}/\sqrt{-M}= \frac{\sqrt{1+a}\pm\sqrt{1-a}}{2},\label{rescalings}
\end{align}
with $a^2\leq 1$. Then, omitting the tildes hereafter, the geodesic equations read
\begin{align}
\label{tdots}
\dot{t}&=  \frac{E u - b_{+}b_{-} L}{(u +b_{+}^2)(u +b_{-}^2)},\\
\label{thetadots}
\dot{\theta} &= \frac{(u+1 )L +b_{+}b_{-} E}{(u +b_{+}^2)(u +b_{-}^2)}, \\
\label{rdotgens}
\frac{\dot{u}^2}{4(-M)} &= -\varepsilon (u +b_{+}^2)(u +b_{-}^2)+(E^2-L^2) u -L^2  - 2 b_{+}b_{-} E L.
\end{align}

The solutions of these equations contain three integration constants $t_0, r_0$ and $\theta_0$. Due to the invariance under rotations and time translations, $t_0$ and $\theta_0$ can be chosen to vanish without loss of generality.

\section{Radial bounds} 

The equation for the radial motion \eqref{rdotgens} is conveniently written as
\begin{equation} \label{rdotgens2}
\frac{\dot{u}^2}{4(-M)}=-\varepsilon u^2 +B u -C \equiv h(u),
\end{equation}
with 
\begin{equation}
B= E^2-L^2-\varepsilon \quad \mbox{and} \quad C=\varepsilon a^2/4 +L^2+a E L. 
\end{equation}

Geodesics exist in the regions $u \ge 0$ where $h(u)$ is non-negative. With this criterion one can find radial bounds for the different geodesics. It is important to note that for timelike and null geodesics, $E^2 \le L^2$ would imply $B<0$ and $C>0$, which in turn implies $h(u)<0$ for all $u>0$. Hence, $E^2 > L^2$ is a necessary condition for the existence of timelike and null geodesics, although not for spacelike ones. 

In what follows, we analyze the $r$-dependence of the different types of geodesics, first for generic $M<0$ and separately for $M=0$.

\subsection{Null geodesics ($\varepsilon =0$)} 

The existence of null geodesics for non-extremal NSs requires $E^2 > L^2$, and since $E \neq 0$, it is useful to define 
\begin{equation} \label{eta}
\eta=\frac{L}{E},
\end{equation}
which verifies $\eta^2 < 1$. Under this condition on $\eta$ the region where $h(u)$ is non-negative depends on the sign of $C=E^2 \eta(\eta +a)$. In the case $\eta(\eta +a)\ge 0$  null geodesics are allowed for $\eta(\eta+a) / (1-\eta^2)\leq u <\infty$. Alternatively, for $\eta(\eta +a) < 0$ the null geodesics are permitted in the half line $0\leq u <\infty$. Table 3 summarizes the possible ranges of $r$ for null geodesics around a naked singularity.

\begin{table}[ht]
\centering
\caption{Radial bounds for null geodesics with $\eta^2  \ngtr 1$}
\begin{tabular}{c|c}
\hline 
\bf{Range of $\boldsymbol{\eta}$ and $\boldsymbol{a}$}  & \bf{Range of $\boldsymbol{r}$} \\ 
 \hline \hline 
 $\eta(\eta+a)\leq 0$ & $0\leq r^2<\infty$ \\ \hline 
$\eta(\eta+a) > 0$ & $0<-M\ell^2 \displaystyle  \frac{\eta(\eta+a)}{1-\eta^2}\leq r^2<\infty$ \\[1.5mm] \hline
$a^2=1$, $\eta a=-1$ & $r$ constant and arbitrary \\ \hline 
\end{tabular}
\end{table}

Integrating Eq.  \eqref{rdotgens2} with $\varepsilon=0$, we obtain
\begin{equation} \label{ulnull}
u(\lambda)=\frac{ \eta(\eta +a)}{1-\eta^2}  -M E^{2} (1-\eta^2)\lambda^2,
\end{equation}
Note that for $\eta(\eta +a) \le 0$  the minimum of the parabola $u(\lambda)$ given by \eqref{ulnull} is non-positive, which implies that any null geodesic coming from a finite radius reaches $u=0$ at a finite value of  $\lambda$. Hence, these null geodesics have no turning point. Meanwhile, in the case  $\eta(\eta +a) >0$ there is a non-zero turning point at 
\begin{equation}
u_{\text{min}}= \frac{\eta(\eta + a)}{1-\eta^2},
\end{equation} 
as shown in Table 3. 

For the extremal naked singularity ($a^2=1$), the cases $\eta^2 >1$ and $\eta a=1$ are not allowed for null geodesics. Remarkably, for $\eta a =-1$,  Eq. \eqref{rdotgens2} provides the circular null geodesics $u(\lambda)=$ constant, where any radius is permissible.  For $\eta^2 <1$, null geodesics are allowed for $0\leq u<\infty$ if $\eta a <0$ or $\eta=0$, and have a turning point if $\eta a >0$, as shown in Table 3.

\subsection{Timelike geodesics ($\varepsilon >0$)}

Timelike geodesics exist in the regions where the quadratic function $h(u)$ in \eqref{rdotgens2}, defined in the domain $u \ge 0$,   is non-negative.  In cases (a) $C<0$ and (b) $C=0$, $B>0$, the function $h(u)$ is non-negative in the interval $0 \leq u \leq u_{+}$, where
\begin{equation} \label{umax}
 u_{\pm}=\frac{B\pm\sqrt{\Delta}}{2 \varepsilon}, \quad \mbox{with} \quad \Delta= B^2 -4\varepsilon C.
\end{equation}
 Note that under conditions (a) or (b), the discriminant $\Delta$ is always positive. A third case is defined by the condition (c) $C>0, B>0, \Delta >0$. Here $h(u) \ge 0$ in the interval $[u_{-}, u_{+}]$.
On the other hand, timelike geodesics are not possible for the cases $C>0$, $\Delta \leq 0$ or $B \le 0$, $C \ge 0$. 

For the static NS ($a=0$), timelike geodesics with $L\neq 0$ satisfy condition (c), since $C=L^2>0$, and thus do not fall into the singularity. The radial timelike geodesics ($a=0, L=0$) satisfy condition (b).

\begin{table}[ht]
\centering
\caption{Radial bounds for timelike geodesics}
\begin{tabular}{c|c}
\hline
\bf{Cases} & \bf{Range of $\boldsymbol{r}$} \\ \hline \hline
(a), (b) & $0\leq r^2\leq -M\ell^2 u_+$ \\ \hline
(c)  & $-M\ell^2u_-\leq r^2\leq -M\ell^2u_+$\\ \hline
\end{tabular} \label{timelikeTable}
\end{table}

The radial equation \eqref{rdotgens2} for $\varepsilon > 0$ is integrated as
\begin{equation} \label{ulnon}
u(\lambda)=\frac{B+\sqrt{\Delta}\sin (2\sqrt{-M \varepsilon} \lambda)}{2\varepsilon},
\end{equation}
which agrees with the cases (a), (b) and (c) previously discussed. In the extremal NS, the cases  $\eta^2 >1$  or $ \eta a= \pm 1$, are also not allowed for timelike geodesics. For  $\eta^2 <1$ the bounds shown in Table 4 are obtained under the condition $a=\pm 1$. Equation (\ref{ulnon}) also holds for the extremal NS. 

\subsection{Spacelike geodesics ($\varepsilon <0$)} 

For spacelike geodesics, $h(u)$ becomes a convex parabola. In the analysis of radial bounds there are three cases to consider:

(a) For $B\geq 0$ and $C>0$, the geodesics stretch from infinity to a minimum radius $r_{\text{min}}^2=-M\ell^2u_-$. 

(b) For $B\geq 0$ and $C\leq 0$, all geodesics end at the singularity. 

(c) For $B<0$ and $C\geq 0$, once again the geodesics can be in the region $u_- \leq u < \infty$. It can be shown that $B<0$ is incompatible with $C<0$, so we do not consider this case.

Solving Eq. \eqref{rdotgens2} with $\varepsilon <0$ yields
\begin{align} \label{slradialsol}
u(\lambda) = \frac{1}{4(-\varepsilon)}\left(e^{2\sqrt{\varepsilon M}\lambda}+\Delta e^{-2\sqrt{\varepsilon M}\lambda} -2B \right).
\end{align}
It can be verified that this solution satisfies the bounds mentioned earlier as summarized in Table \ref{spacelikebounds}. This table also provides radial bounds for the extremal case $a^2=1$, for which the solution (\ref{slradialsol}) holds as well.
\begin{table}[ht]
\centering
\caption{Radial bounds for spacelike geodesics} \label{spacelikebounds}
\begin{tabular}{c|c}
\hline 
\bf{Cases} & \bf{Range of $\boldsymbol{r}$} \\ \hline \hline 
(b) & $0\leq r^2<\infty$ \\ \hline
(a), (c) & $-M\ell^2u_-\leq r^2<\infty$
\\  \hline
\end{tabular}
\end{table}

A qualitative summary of the  results for radial bounds is that null and spacelike geodesics approach the NS from infinity, and either fall into the singularity or wind around and go back to infinity (with the exception of the extremal case for which a null geodesic with $\eta a=-1$ has a circular orbit). Meanwhile timelike geodesics either orbit continually bounded between two radii, or fall into the NS.  

\subsection{Geodesics on the massless BTZ geometry} \label{Massless BTZ Geodesics}
 
The geodesic equations for the massless BTZ spacetime ($M=J=0$), written in terms of the original variables\footnote{In this case the rescalings (\ref{rescalings}) are no longer valid.} of Eqs. \eqref{tdot}$-$\eqref{rdotgen}, are given by
\begin{equation}
\label{geoEqmBTZ}
\dot{t}= \frac{E \ell^2}{r^2},\quad \dot{\theta} = \frac{L}{r^2}, \quad \dot{r}^2 = -\varepsilon \frac{r^2}{\ell^2}+E^2-\frac{L^2}{\ell^2}.
\end{equation}
The integration of these equations is straightforward and Table 6 summarizes the solutions of the radial equation and bounds. Note that $L=0$ provides radial geodesics and the case $ E^2<L^2/\ell^2$ is not allowed for null and timelike geodesics. Moreover, the bounds match those for $M \neq 0$ in the limit $M\to 0$, c.f. \cite{Cruz1994}.

\begin{table}[ht]
\footnotesize
\centering
\caption{Radial bounds and solutions for geodesics of the massless BTZ spacetime. The signs $\pm$ refer to outgoing/ingoing geodesics. The time component can be obtained as $t(\lambda)=E \ell^2 \theta(\lambda)/L$. } \label{masslessbounds}
\begin{tabular}{c|c|c}
\hline
\bf{Case} & \bf{Range of $\boldsymbol{r}$} &  $\boldsymbol{r(\lambda)}, \boldsymbol{\theta(\lambda)}$\\ \hline \hline 
$\varepsilon=0$, $E^2>L^2/\ell^2$ & $0  \leq r  <\infty $ & $ \begin{array}{c} r= \pm\sqrt{E^2-L^2/\ell^2}\lambda +r_0 
  \\ \theta=\frac{ \mp L}{\sqrt{E^2-L^2/\ell^2} \left(\pm \sqrt{E^2-L^2/\ell^2}\lambda +r_0\right)}   \end{array} $ \\ \hline 
 $\varepsilon=0$, $E^2=L^2/\ell^2$ & $r$ constant and arbitrary &  $ \begin{array}{c}  r= r_0\\
 \theta= \frac{L}{r_0^2} \lambda   \end{array} $ \\
\hline
$\varepsilon>0$, $E^2>L^2/\ell^2$ & $0 \leq r\leq \ell \displaystyle\sqrt{\frac{E^2-L^2/\ell^2}{\varepsilon}}$  & $ \begin{array}{c}  r= \ell\sqrt{\frac{E^2-L^2/\ell^2}{\varepsilon}} \sin \left( \frac{\sqrt{\varepsilon}}{\ell}\lambda \right) \\
\theta= -\frac{\ell L \sqrt{\varepsilon } }{E^2 \ell^2-L^2} \cot \left(\frac{ \sqrt{\varepsilon }}{\ell}\lambda \right)  \end{array} $ \\
\hline 
$\varepsilon<0$, $E^2 > L^2/\ell^2$ & $0 \leq r<\infty$ & $ \begin{array}{c} r= \ell\sqrt{\frac{E^2-L^2/\ell^2}{-\varepsilon}} \sinh \left(\pm\frac{\sqrt{-\varepsilon}}{\ell}\lambda \right)\\
\theta= -\frac{\ell L \sqrt{-\varepsilon } }{E^2 \ell^2-L^2}  \coth \left(\frac{ \sqrt{-\varepsilon }}{\ell}\lambda \right)   \end{array} $ \\
\hline 
$\varepsilon<0$, $E^2 = L^2/\ell^2$ & $0 < r<\infty$ & $ \begin{array}{c}  r= r_0 \, e^{ \pm\frac{\sqrt{-\varepsilon}}{\ell}\lambda} \\
\theta= \mp \frac{\ell L }{2 r_0^2 \sqrt{-\varepsilon }} e^{\mp\frac{2  \sqrt{-\varepsilon }}{\ell} \lambda } \end{array} $ \\
\hline
$\varepsilon<0$, $E^2<L^2/\ell^2$ & $\ell \displaystyle \sqrt{\frac{E^2-L^2/\ell^2}{\varepsilon}} \leq r<\infty$ & $ \begin{array}{c} r= \ell\sqrt{\frac{E^2-L^2/\ell^2}{-\varepsilon}} \cosh \left( \frac{\sqrt{-\varepsilon}}{\ell}\lambda \right)\\ 
 \theta= \frac{\ell L \sqrt{-\varepsilon } }{E^2 \ell^2-L^2} \tanh \left(\frac{ \sqrt{-\varepsilon }}{l} \lambda\right) \end{array} $ \\ \hline
\end{tabular}
\end{table}

In the null case ($\varepsilon=0$), for $E^2>L^2/\ell^2$, we obtain
\begin{align}
    r(\lambda)=\pm\sqrt{E^2-L^2/\ell^2}\lambda +r_0,
\end{align} 
where $r_0>0$ is an arbitrary integration constant. The range of the affine parameter is  $0\le\lambda<\infty$ for the upper sign, and $-\infty < \lambda\le r_0/\sqrt{E^2-L^2/\ell^2}$ for the lower sign. For $E^2=L^2/\ell^2$, $r(\lambda)$ is constant (circular orbit).

For the timelike case ($\varepsilon>0$), the solution of the radial equation is given by  
\begin{equation} \label{rtlmBTZ}
r(\lambda)= \ell\sqrt{\frac{E^2-L^2/\ell^2}{\varepsilon}} \sin \left( \frac{\sqrt{\varepsilon}}{\ell}\lambda \right),
\end{equation}
with $0<\lambda<\frac{\ell}{\sqrt{\varepsilon}}\pi$, i.e., $\lambda$ is bounded. 

Finally, for spacelike geodesics ($\varepsilon<0$), the solution of the radial equation is
\begin{equation} \label{rslmBTZ}
r(\lambda)= \begin{cases} \displaystyle \ell\sqrt{\frac{E^2-L^2/\ell^2}{-\varepsilon}} \sinh \left(\pm\frac{\sqrt{-\varepsilon}}{\ell}\lambda \right), & \mbox{if } E^2 > L^2/\ell^2 \\[2mm]
 r_0 \, \displaystyle e^{ \pm\frac{\sqrt{-\varepsilon}}{\ell}\lambda}, & \mbox{if } E^2 = L^2/\ell^2 \\[2mm] \displaystyle \ell\sqrt{\frac{E^2-L^2/\ell^2}{-\varepsilon}} \cosh \left( \frac{\sqrt{-\varepsilon}}{\ell}\lambda \right), & \mbox{if } E^2 < L^2/\ell^2 \end{cases}
\end{equation}
where the upper (lower) sign corresponds to outgoing (ingoing) geodesics. For $E^2\geq L^2/\ell^2$, the range of the affine parameter is $0 \le \lambda < \infty$ for outgoing geodesics, and $-\infty  <\lambda \le 0$ for the incoming case. Otherwise, if $E^2 < L^2/\ell^2$, the affine parameter can take all real values.

\section{Orbits in the $r$-$\theta$ plane} \label{Polar orbits} 

Now, we analize the orbit equation $r(\theta)$, which can be obtained from \eqref{thetadots} and \eqref{rdotgens}. Once again we begin with the non-extremal case.  
Integration of the orbit equation yields 
\begin{equation} \label{orbit2}
2\sqrt{-M}\theta=A_+I_+(u) + A_-I_-(u),
\end{equation}
with 
\begin{equation}
\quad A_{\pm} = \mp \frac{b_{\mp}}{b_+^2-b_-^2}
\end{equation}
and 
\begin{equation} \label{Ipm}
I_{\pm}= \begin{cases} \displaystyle  \arctan \frac{D_{\pm}(u+b_{\pm}^2)-2F_{\pm}^2}{2 F_{\pm}\sqrt{-\varepsilon u^2 +B u -C}}, & \mbox{if } \varepsilon \neq 0 \\[4mm]
 \displaystyle 2  \arctan \frac{\sqrt{(1-\eta^2) u -\eta (\eta+a) }}{ b_{\pm} +\eta b_{\mp}}, & \mbox{if }\varepsilon = 0,  \end{cases}
\end{equation}
where $ 
D_{\pm}= \pm \varepsilon(b_+^2-b_-^2)+E^2-L^2$ and $F_{\pm}=b_{\pm}E+b_{\mp}L\neq 0$. 

For null and timelike geodesics, $E^2 > L^2$ holds, so $F_+ \neq 0$. Then, in the sub-case $F_+ \neq 0, F_-=0$, Eq. \eqref{orbit2} reduces to $2\sqrt{-M}\theta = A_+I_+$. On the other hand, $F_+=0,F_-\neq 0$ {\it is} allowed for a spacelike geodesic, and in this case $2\sqrt{-M}\theta = A_-I_-$.

In the static case ($a=0$), Eq. \eqref{thetadots} gives the radial geodesic $\theta= \textrm{constant}$ for $L=0$. Note that there are no radial geodesics if $a\neq 0$, i. e., rotating NSs always produce dragging.

Null geodesics with $L\neq 0$ have the simple expression 
\begin{equation} \label{orbitnullstatic}
r^2(\theta)=\frac{-M\ell^2\eta^2}{(1-\eta^2) \cos^2( \sqrt{-M} \theta)},
\end{equation}
which has a turning point at $r_{\text{min}}= \ell \sqrt{\frac{-M \eta^2}{1-\eta^2}}$. For timelike and spacelike geodesics,
\begin{equation} \label{orbittlstatic}
r^2(\theta)=\frac{-2M\ell^2 L^2}{B +\sqrt{B^2-4\varepsilon L^2}\cos (2\sqrt{-M}\theta)},
\end{equation}
in agreement with the radial bounds shown for  condition (c) in Table 4. In the limit $\varepsilon \to 0$, Eq. \eqref{orbittlstatic} matches the orbit for the static null geodesics \eqref{orbitnullstatic}. Note that the integration constant here has been chosen as $\theta_0\neq 0$, unlike in (\ref{orbit2}), to make the expression \eqref{orbittlstatic} simpler. 

\begin{figure}[ht!]
\begin{subfigure}{0.495\textwidth}
\centering
\includegraphics[height=2.7in]{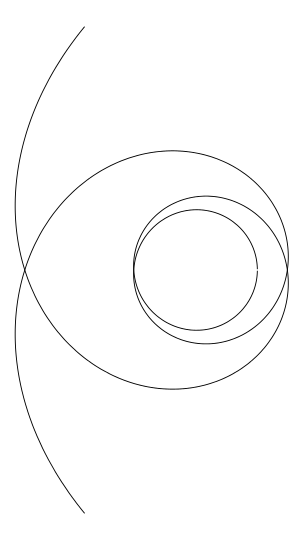}
\caption{$\eta a>0$} \label{encouraged}
\end{subfigure}
\begin{subfigure}{0.49\textwidth}
\centering
\includegraphics[height=2.7in]{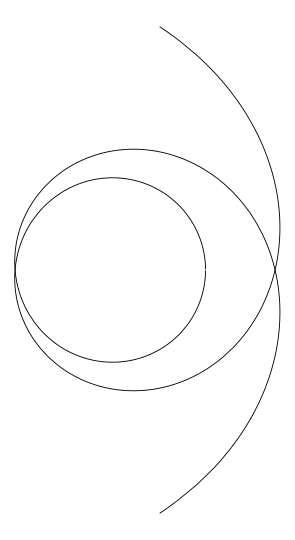}
\caption{$\eta a <0$} \label{diminshed}
\end{subfigure}
\caption{Null geodesics with equal $|\eta a|$}
\end{figure}

We include plots that help visualize the different geodesics. Note in particular the possibility of null geodesics to reach infinity. Additionally, the winding of geodesics near the NS can either be increased or decreased, depending on the magnitude of $L$ and its sign relative to $J$. This dragging produced by the rotation of the NS, can be seen in figures \ref{encouraged} and \ref{diminshed}. 

As we adjust different parameters (in particular, $M$), the number of times geodesics wind around the NS and intersect themselves changes. This winding phenomenon is studied in detail in the following section.

\begin{figure}[ht!] 
\begin{subfigure}{0.3\textwidth}
\centering
\includegraphics[height=1.8in]{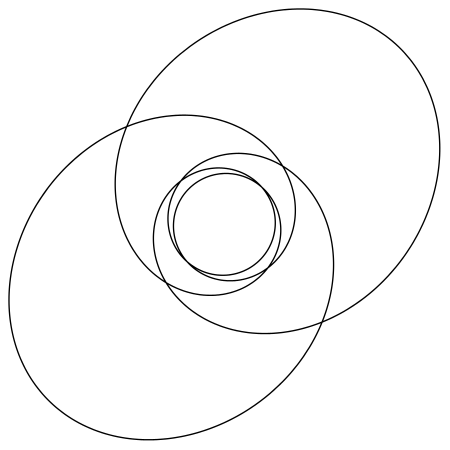}
\caption{$\sqrt{-M}=1/5$} \label{timelikemass1}
\end{subfigure}
\begin{subfigure}{0.3\textwidth}
\centering
\includegraphics[height=1.8in]{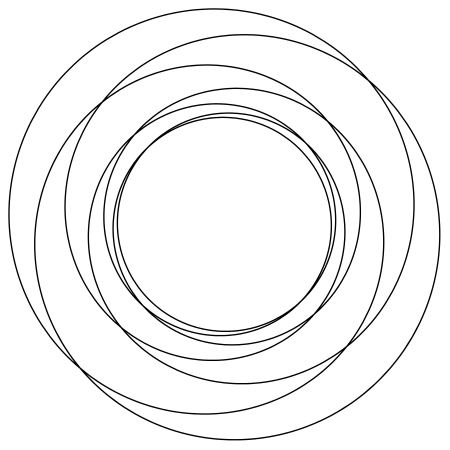}
\caption{$\sqrt{-M}=1/17$} 
\end{subfigure}
\begin{subfigure}{0.3\textwidth}
\centering
\includegraphics[height=1.8in]{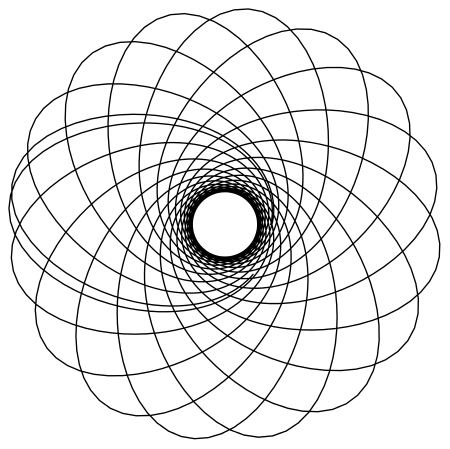}
\caption{$\sqrt{-M}=1/\sqrt{15}$} \label{dramatic}
\end{subfigure}
\caption{Timelike geodesics} \label{timelikefig}
\end{figure}

In agreement with the solutions discussed in the previous section, timelike geodesics follow bounded orbits around the NS. For $C>0$ and rational values of $\sqrt{-M}$, the orbits are closed. As the denominator of $\sqrt{-M}$ (expressed as an irreducible fraction) grows, timelike geodesics take more winds to close. Figure \ref{dramatic} is an  example of the failure of timelike geodesics to close when $\sqrt{-M}$ is irrational. In this figure $\lambda$ is not allowed to cover its entire range, for otherwise nothing would be visible. 

The extremal naked singularity produces different solutions for orbits.
For null geodesics the orbit equation is defined only if $\eta^2 <1$ and the solution\st{s are} \bc{is}
\begin{align}\label{orbitnullext}
 2\,\sgn(E)\sqrt{-M}\theta&=  \frac{\sqrt{(1-\eta^2) u -\eta(\eta\pm1)}}{(\eta \pm 1)(u+1/2)}   \nonumber \\ &+
 \sqrt{2}  \arctan \frac{\sqrt{2}\sqrt{(1-\eta^2) u -\eta(\eta\pm1)}}{(\eta\pm 1)}.
\end{align}

As for timelike and spacelike geodesics, we have
\begin{align} \label{orbittlext}
&2\sgn(E)\sqrt{-M}\theta=  \frac{\sqrt{-\varepsilon (u+1/2)^2/E^2 +(1-\eta^2) u -\eta(\eta\pm1)}}{(\eta \pm 1)(u+1/2)}   \nonumber \\ &+
  \frac{ 1}{\sqrt{2}}  \arctan \frac{(1-\eta^2)(u+1/2)-(1\pm \eta)^2}{\sqrt{2}(\eta\pm1)\sqrt{-\varepsilon (u+1/2)^2/E^2 +(1-\eta^2) u -\eta(\eta\pm1)}}.
\end{align}

We note that $\sgn(E)$ can always be chosen to be positive for the null and timelike cases, as will be discussed in Section 7.

\subsection{Orbits for the massless BTZ geometry} 

Finally we consider  orbits in the massless BTZ spacetime. 
The non-radial ($L\neq 0$) and non-circular ($E^2\neq L^2/\ell^2$) orbits for all geodesics are given by
\begin{equation} \label{orbitmBTZ}
 r^2(\theta)= \frac{L^2}{(E^2 -L^2/\ell^2)\theta^2+\frac{\varepsilon L^2}{E^2 \ell^2 -L^2}}.
 \end{equation}
where $\varepsilon$ can be set as $\pm 1$, or $0$.
 
For null or spacelike geodesics and $E^2>L^2/\ell^2$, this equation describes a spiral connecting $r=0$ with $r=\infty$. Additionally, there is a reflected spiral due to the symmetry $\theta \to -\theta$ of Eq. \eqref{orbitmBTZ} (not shown in Fig. 3). 
 
\begin{figure}[ht!] 
\begin{subfigure}{0.3\textwidth}
\centering
\includegraphics[height=1.9in]{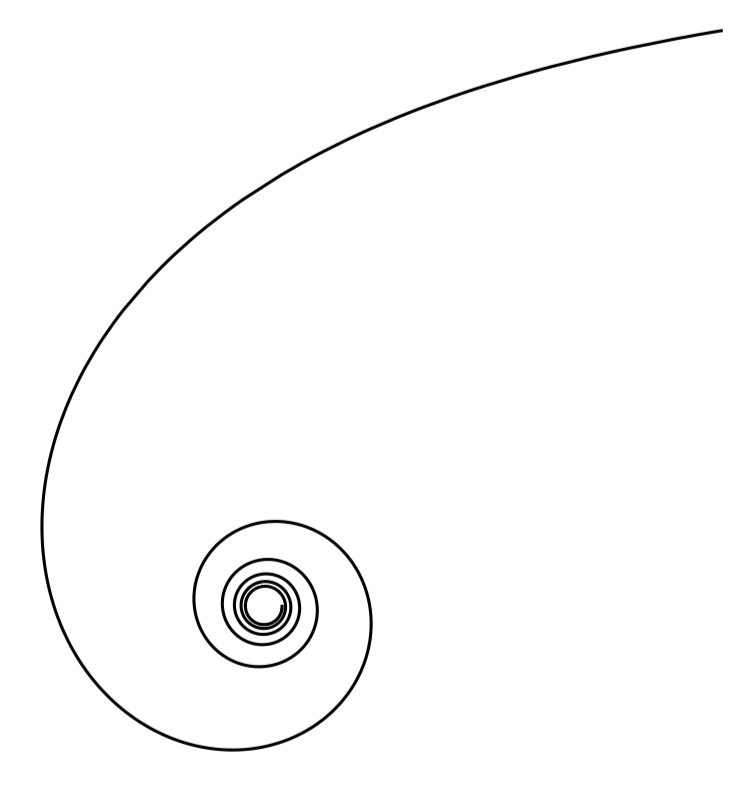}
\caption{Null} \label{nullmass}
\end{subfigure}
\begin{subfigure}{0.3\textwidth}
\centering
\includegraphics[height=1.9in, angle=0]{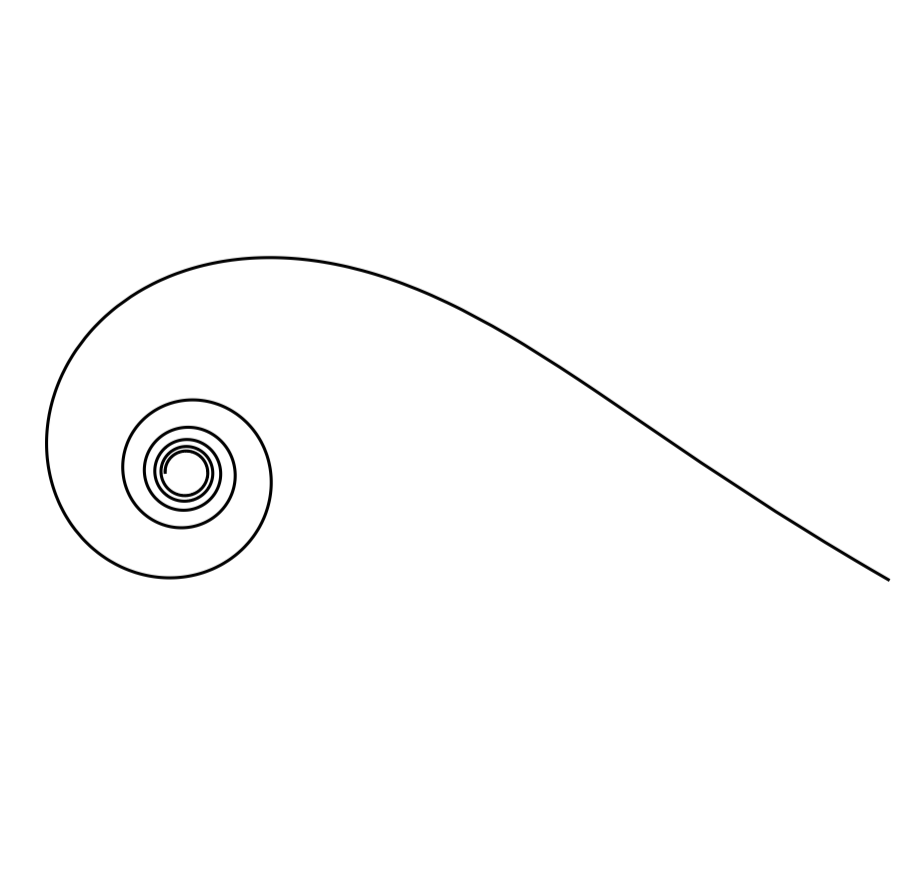}
\caption{Spacelike} \label{spacelikemass}
\end{subfigure}
\begin{subfigure}{0.29\textwidth}
\centering
\includegraphics[height=1.9in]{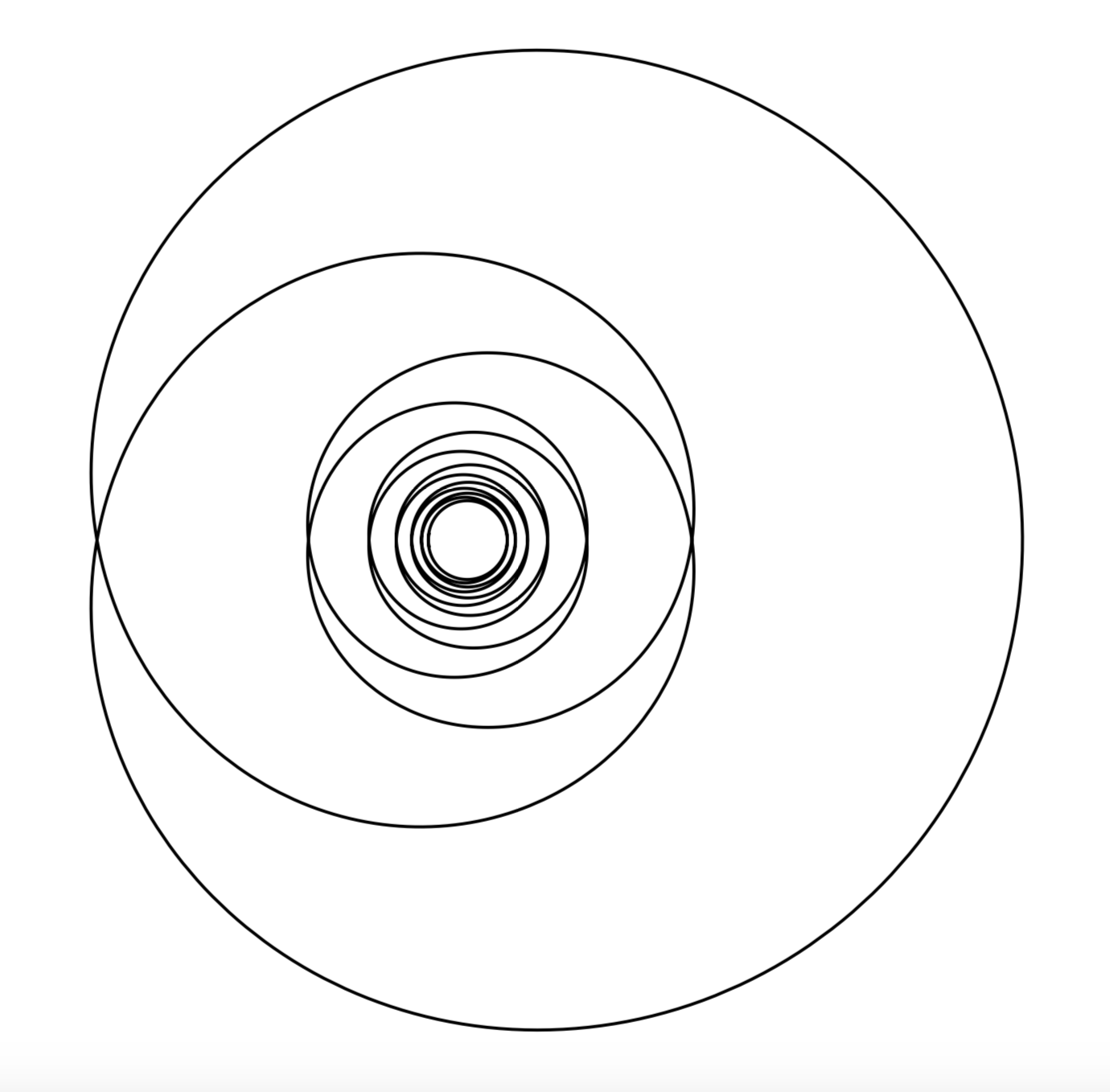}
\caption{Timelike} \label{timelikemass2}
\end{subfigure}
\caption{Geodesics of the massless BTZ spacetime for $E^2>L^2/\ell^2$. The ingoing null and spacelike geodesics spiral into $r=0$, while the outgoing ones spiral away from the origin to infinity. Timelike geodesics are spirals bounded by $0<r<r_{\textrm{max}}$ as shown in Table 6.} \label{allmassless}
\end{figure}

 Timelike geodesics --which require $E^2>L^2/\ell^2$--  are spirals connecting the origin for $\theta \to -\infty$ to a maximum finite radius $r_{\textrm{max}}=\sqrt{(E^2\ell^2 - L^2)/\varepsilon}$ at $\theta=0$. Then, after an infinite number of turns they return to the origin for $\theta \to \infty$. Note that as shown in Eq. (\ref{rtlmBTZ}), the entire geodesic is covered by an affine parameter $\lambda$ that is bounded for timelike geodesics of the massless BTZ spacetime. Although the radial motion for a timelike geodesic (\ref{rtlmBTZ}) is  sinusoidal, it does not describe an oscillation near the singularity.\footnote{This point was incorrectly interpreted in \cite{strings}.} In Figure \ref{allmassless} we illustrate each type of geodesic. Globally they are quite different, though close to the singularity all geodesics simply look like the aforementioned spirals. 

The case $E^2<L^2/\ell^2$ for spacelike geodesics is also considered in (\ref{orbitmBTZ}), but here the geodesics do not reach the singularity. Instead they have a finite minimum radius as indicated in Sec. \ref{Massless BTZ Geodesics}. In these spacelike geodesics $\theta(\lambda)$ is bounded as $-\theta_\infty<\theta(\lambda)<\theta_\infty$, where
\begin{equation} \label{anginf}
\theta_\infty= \frac{\ell L }{L^2-E^2 \ell^2}
\end{equation}
is the angle for $r \to \infty$. If $\theta_\infty \le \pi$, the geodesic comes from infinity and goes back to infinity without self-intersections. For $\theta_\infty > \pi$ the geodesic wind around the singularity a number of times before going back to infinity. This behavior is illustrated with two examples in Figure 4.
 
\begin{figure}[ht!] 
\begin{subfigure}{0.5\textwidth}
\centering
\includegraphics[height=2in]{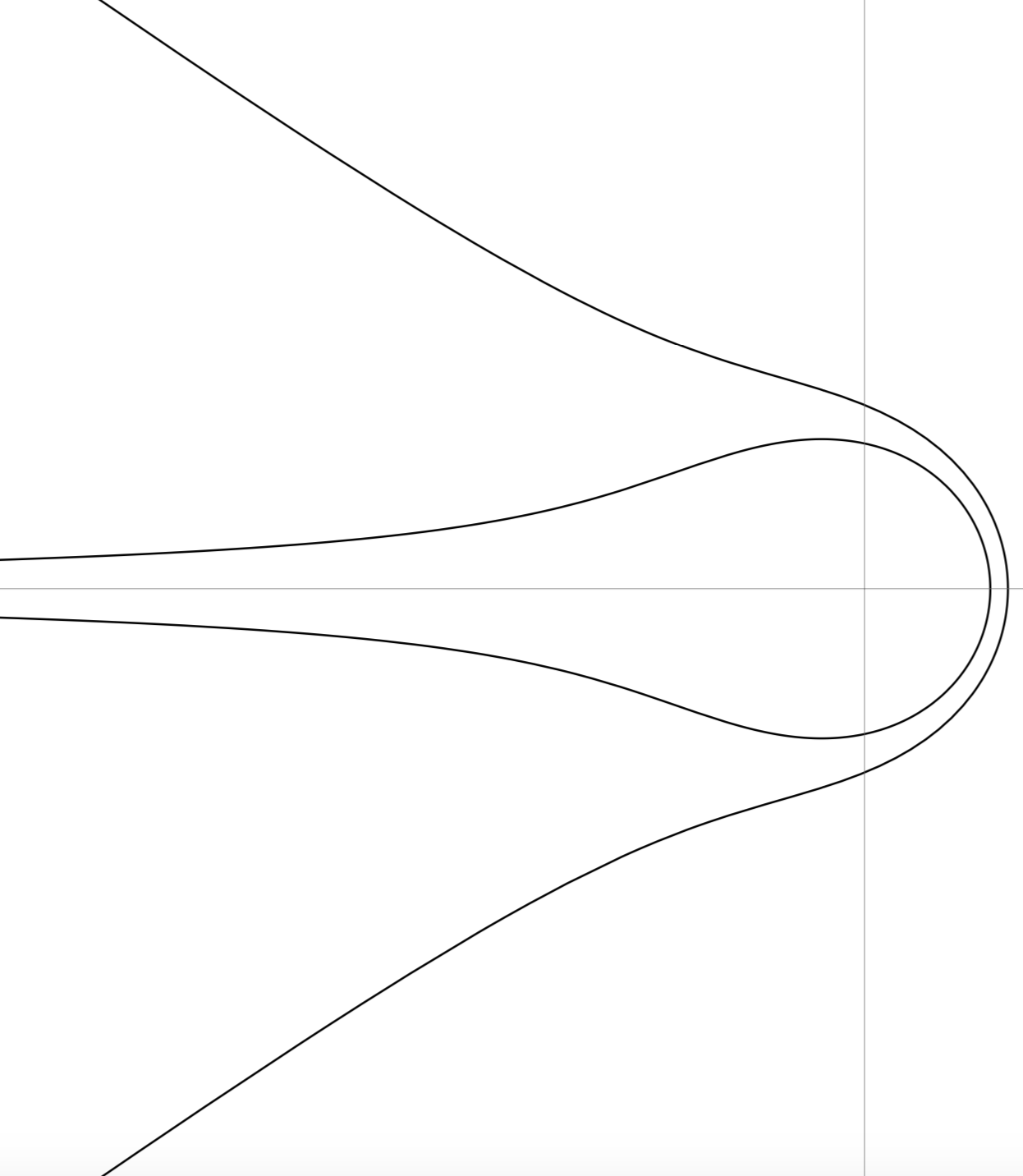}
\caption{$\theta_\infty=0.8 \pi$ and $\theta_\infty=\pi$} \label{spacelikemass41}
\end{subfigure}
\begin{subfigure}{0.5\textwidth}
\centering
\includegraphics[height=2in]{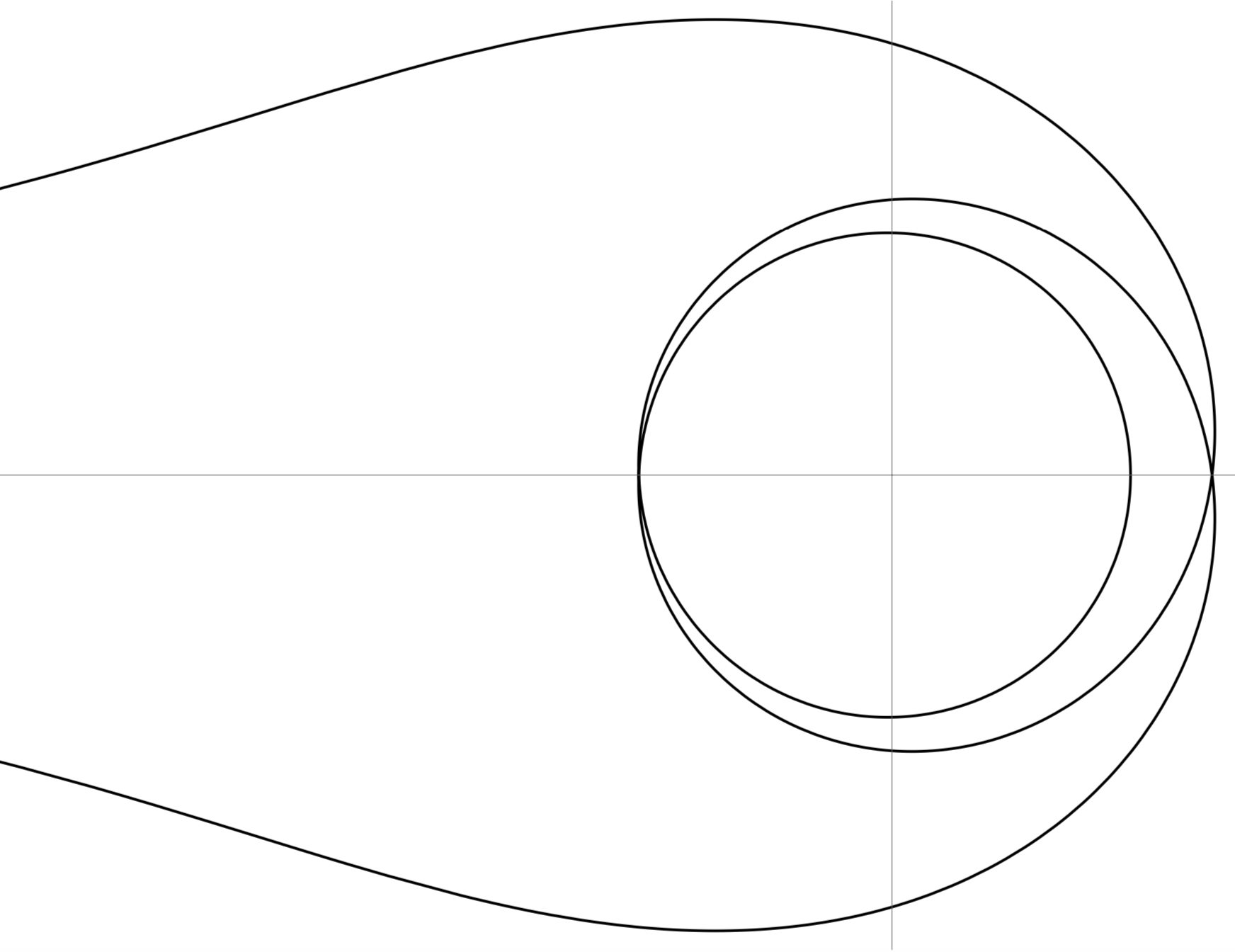}
\caption{$\theta_\infty=3 \pi$} \label{spacelikemass43}
\end{subfigure}
\caption{Spacelike geodesics of the massless BTZ spacetime with $E^2<L^2/\ell^2$.} \label{splmassless}
\end{figure}

\section{Self-intersections}  
Null and spacelike geodesics that have a turning point intersect themselves. These geodesics start at infinity and wind around the singularity a finite number of turns, reach the turning point $r_{\text{min}}$ and go back to infinity after repeating the same number of turns. The number of self-intersections is the integer number of times that $2\pi$ is contained in the angle swept by the geodesic as $r$ goes from $r=\infty$ to $r=r_{\text{min}}$ and back to $r=\infty$. First, we count self-intersections for $M\neq 0$. 

Let us consider null geodesics with a turning point $r_{\text{min}}$, which require  $\eta^2 <1$  and $\eta(\eta+ a) >0$. Setting the angle at $r_{\text{min}}$ equal to 0, then at $r=\infty$ it is
\begin{equation} \label{tmax}
|\theta(\infty)|=\frac{\pi}{2 \sqrt{-M}(b_+^2-b_-^2)}|  b_{+}\sgn(b_- +\eta b_+)- b_{-}\sgn( b_+ +\eta b_-)|.
\end{equation}
Since $b_+> b_-$ and $\eta^2<1$, we have $ \sgn( b_+ +\eta b_-)=1$. Moreover, if $\eta(\eta+a)>0$ we have  $\sgn(b_- +\eta b_+)=\sgn(\eta)$ and therefore,
\begin{equation} \label{t1}
\mathcal{N}:=\left|\frac{\theta(\infty)}{\pi}\right|=\frac{1}{2\sqrt{-M+\sgn(\eta)J/\ell}}.
\end{equation}
The number of self-intersections ($N$) is the integer part of $\mathcal{N}$, except for $\mathcal{N} \in \mathbb{Z}$, in which case one must subtract 1 (which corresponds to a self-intersection at $r=\infty$). Hence, this number is given by 
\begin{equation} \label{techo}
N_{\pm}=\left\lceil\frac{1}{2\sqrt{-M +\sgn(\eta) J/\ell}}\right\rceil-1,
\end{equation}
where the ceiling function $\lceil x\rceil$ is the least integer greater than or equal to $x$, and  $\pm = \sgn(\eta)$. The difference between $N_+$ and $N_-$ for  a given sign of $J$ is due to the fact that the rotating background breaks the clockwise/counter-clockwise symmetry of the null geodesic. Once again, we see that the rotation of the NS can either increase or decrease the winding of the geodesic. 

For the extremal case, self-intersections occur only if $\sgn(\eta)a=1$, then we get
\begin{align}
    N = \left\lceil \frac{1}{2\sqrt{-2M}}\right\rceil -1.
\end{align}

 We can also obtain the winding number for spacelike geodesics, which start at infinity and reach a minimum radius, as $|\theta(\infty)-\theta(r_{\min})|/\pi$. Assuming $F_{\pm}\neq 0$ in (\ref{Ipm}), the angle at the turning point is given by
\begin{align}
\theta(r_{\min}) = \frac{\pi}{4\sqrt{-M}(b_+^2-b_-^2)}[-b_-\sgn(F_+(D_+(u_{\text{min}}+b_+^2)-2F_+^2))\\+b_+\sgn(F_-(D_-(u_{\text{min}}+b_-^2)-2F_-^2))]. \nonumber
\end{align}
From this it is relatively straightforward to show that 
\begin{align} \label{thetamin}
\theta(r_{\min}) = \frac{\pi}{4\sqrt{-M}(b_+^2-b_-^2)}[b_-\sgn(F_+)-b_+\sgn(F_-)],
\end{align}
while at $r=\infty$ one finds,
\begin{align} \label{thetainf}
\theta(\infty) =\frac{1}{2\sqrt{-M}(b_+^2-b_-^2)}\left(-b_- \arctan \frac{D_+}{2F_+\sqrt{-\varepsilon}}+b_+ \arctan \frac{D_-}{2F_-\sqrt{-\varepsilon}}\right)
\end{align}
As explained in Section \ref{Polar orbits}, if $F_+=0$ only the second terms in \eqref{thetamin} and \eqref{thetainf} must be considered, while if $F_-=0$ just the first ones appear in those expressions. 

Self-intersections are also present in timelike orbits. They occur in case (c) of Table \ref{timelikeTable}, where timelike geodesics are bounded by two radii. However, in general orbits do not close which leads to an infinite number of intersections (see for instance Figure \ref{dramatic}).

For the massless BTZ spacetime, self-intersections occur for time- and spacelike geodesics. Initially outgoing timelike geodesics with $E^2>L^2/\ell^2$ (see Figure 3(c)) can have an arbitrarily large number of self-intersections depending on the initial conditions;  for spacelike geodesics with $E^2<L^2/\ell^2$ (see Figure 4(b)), using $\theta(\infty)$ in Eq. \eqref{anginf}, the number of the self-intersections is found to be
\begin{align}
    N = \left\lceil \frac{\ell L }{(L^2-E^2 \ell^2)\pi}\right\rceil -1.
\end{align}

It is worth noting that the number of self-intersections for null geodesics depends only on the {\it spacetime quantities} $M$ and $J$, while for spacelike and timelike geodesics, the number of self-intersections also depends on the constants of the orbit $E$ and $L$. 

\section{Additional features}  
\subsection{The time coordinate $t$} 
So far we have not discussed the behavior of the time coordinate for geodesics in the NS spacetimes. We note, firstly, that the equation of motion (\ref{tdots})  for $t$  closely resembles equation (\ref{thetadots}) for $\theta$. Indeed, integrating (\ref{tdots}) gives a very similar expression to (\ref{orbit2}),
\begin{align}
    2\sqrt{-M}(t-t_0) = \tilde{A}_+ I_+ + \tilde{A}_-I_-,
\end{align}
where 
\begin{align}
    \tilde{A}_{\pm}=A_{\mp}=
     \pm \frac{b_{\pm}}{b_+^2-b_-^2}, 
\end{align}
and $I_{\pm}$ are given by (\ref{Ipm}). 

Another aspect of the time coordinate worth noting is the sign of $\dot{t}(\lambda)$ throughout a trajectory. We can see from the geodesic equation (\ref{tdots}) that $\dot{t}$ could change sign at $u_c\equiv \eta b_+b_- = \eta a/2>0$. However, this does not occur for timelike or null geodesics, which can be seen as follows:

i) Null geodesics are either radial or never reach the singularity and have a turning point at $u_{\text{min}}>u_c$. Hence, massless particles never reach the $u_c$ point, and the sign of $\dot{t}$ for their geodesics is constant.

ii) For timelike geodesics it can be shown that $h(u_c)<0$ in Eq. (\ref{rdotgens2}), and therefore $u_c$ lies outside of the allowed range for $u$.

Additionally, from Eq. (\ref{tdots}) note that as $u\to \infty$, $\sgn(\dot{t})=\sgn(E)$ and thereore, without loss of generality one can always choose $E>0$ and $\lambda$ so that $t(\lambda)$ is monotonically increasing for timelike and null geodesics. This means that there are no closed timelike geodesics in a NS spacetime in AdS$_3$ \cite{BTZ2}. On the other hand, for spacelike geodesics no such restrictions exist and both $\dot{t}$ and $E$ can have any sign, and could even vanish for the entire geodesic.

\subsection{Lengths of spacelike geodesics} 

The  lengths of spacelike geodesics have been relevant for some time now due to the Ryu-Takayanagi (RT) prescription for computing entanglement entropy from the area of minimal surfaces \cite{RT}. The relevant geodesic is a purely spatial curve $\dot{t}=0$ and therefore we consider the spacelike geodesics with $E=0$ and $J=0$. The length of a geodesic arc is given by  $ds=\sqrt{-\varepsilon}d\lambda$. Then, the arc length of a geodesic path that sweeps out an angle $0<\alpha<\pi/2$ in going from $r=\infty$ back to $r=\infty$ is, 
\begin{align}
    \lambda = 2 \int_0^{\alpha} \frac{d\theta}{\dot{\theta}},
\end{align}
where $\theta=0$ corresponds to the minimum radius.
The integration yields\footnote{In the original variables, save for a re-scaling of $L$ by $\ell$. Here, $B=E^2-L^2-M$, with $\varepsilon=-1$.},
\begin{align} \label{lambda}
    \lambda = \ell \log \left(\frac{L^2-M - (L^2+M)\cos2\sqrt{-M}\theta + 2L\sqrt{-M} \sin2\sqrt{-M}\theta}{(L^2-M)\cos2\sqrt{-M}\theta- (L^2+M)}  \right) \Big|^{\alpha}_0
\end{align}
where $\alpha$ satisfies 
\begin{equation} \label{cos a}
\cos2\sqrt{-M}\alpha = (L^2+M)/(L^2-M).
\end{equation} 
As could be expected, expression (\ref{lambda}) diverges. To give physical meaning to this length it is necessary to regularize it by subtracting off another divergent length. In this case, a reasonable choice is to compare with the corresponding geodesic length in AdS$_3$ $(M=-1)$, which in a sense corresponds to the vacuum geometry. A problem is that there can be different meanings for ``corresponding geodesics'' between different spacetimes. One can choose geodesics that have the same parameters $L$ and $E$, or that sweep the same angle $\alpha$. 
These two options lead to different results. Indeed, comparing geodesics of the same $L$ (and $E=0$) leads to 
\begin{align} \label{lambda-L}
    \Delta \lambda|_{L} \equiv \lambda|_{L}- \lambda_{\text{AdS}}|_{L} = 0,
\end{align}
i.e., the length of spacelike geodesics with $E=0$ and the same angular momentum $L$ are equal in a NS spacetime as in AdS. On the other hand, comparing geodesics that sweep the same angle $\alpha$ gives
\begin{align}\label{lambda-alfa}
    \Delta \lambda |_{\alpha} \equiv \lambda|_{\alpha} - \lambda_{\text{AdS}}|_{\alpha}=  \ell \log \left(\frac{1}{\sqrt{-M}}\frac{\sin2\sqrt{-M}\alpha}{\sin2\alpha}\right).
\end{align}
To arrive at this expression, the regularization leading to \eqref{lambda-L} is obtained as
\begin{align}
    \Delta\lambda|_{L}=\lim_{\theta\to\alpha(L)} \lambda|_{L} - \lim_{\theta\to\tilde{\alpha}(L)}  \lambda_{\text{AdS}}|_{L}
\end{align}
where  $\alpha$ is defined by Eq. \eqref{cos a},  and $\tilde{\alpha}$ is defined analogously but for $M=-1$. Thus, in order to subtract the limits, one must make an appropriate change of variables first. The two expressions (\ref{lambda-L}) and (\ref{lambda-alfa}) could be thought of as arising in different thermodynamic ensembles: in the first case the angular momentum is held fixed, while in the second its canonical conjugate (the angle $\alpha$) is fixed. 

Our results for $\Delta\lambda$ differ from those found by other authors. For instance, the expression for $\Delta \lambda $ in Eq. (2.5) of \cite{lengthsRT} (with $n=1/\sqrt{-M}$)  in our notation reads 
\begin{align} \label{lambda17}
    \Delta\lambda = 2\ell \log \left(\frac{2\ell}{\mu}\sin\sqrt{-M}\alpha \right),
\end{align}
where $\mu$ is a regulator and the discrepancy may be attributed to the different regularization prescription. 

The expression (\ref{lambda-alfa}) naturally vanishes for $M=-1$ but yields a finite result in the limit $M\to 0^-$, which would not match the vanishing entropy of a massless BTZ black hole. In contrast with this, (\ref{lambda17}) diverges for $M\to 0^-$ but gives a non vanishing result for AdS, which means that that regularization corresponds to comparing with a different geometry. The result (\ref{lambda-L}), on the other hand, seems reasonable if the NS geometry is viewed as a point particle that has no horizon and therefore no Hawking temperature and no entropy. There is an additional difficulty with the interpretation of (\ref{lambda-alfa}): Since the NS geometry is obtained by removing an angular sector from AdS, this renders uncertain the idea of comparing lengths with {\it{the same opening angle}} in the two geometries. For instance, it is not clear what happens if the removed angle is larger than  $2(\pi-\alpha)$.

\section{Conclusions and discussion}  
We have investigated the geodesic structure of NS in three-dimensional anti-de Sitter spacetime. We found that null and spacelike geodesics have a finite number of self-intersections. This occurs when they do not reach the singularity, but rather start and end at infinity. For null geodesics the number of self-intersections is a simple expression that depends only on properties of the spacetime ($M$ and $J$), independently of the values $E,L$ of the particular geodesic. On the other hand, for spacelike geodesics the number of self-intersections depends on the constants $E,L$ as well. Timelike geodesics have bounded orbits and also intersect themselves, but this behavior is more complicated, as shown in Figure 2. For example, for the static NS the orbits do not close unless $1/\sqrt{-M}$ is a rational number. When they do not close, they have an infinite number of self-intersections.

It is interesting to point out differences and similarities with the $M>0$ case. For the black hole, the dependence of $\theta$ on $r$ is logarithmic and no self-intersections are present. For the black hole, massive particles always fall into the singularity, while for the NS this does not happen in case (c) of Table \ref{timelikeTable} (see also Figure \ref{timelikefig}). Null geodesics can escape the singularity for both black holes and NS, but radial bounds are complementary: if a null geodesic has parameters $E, L$ such that it reaches the NS, a geodesic with those same parameters will not reach the black hole singularity. Conversely, a null geodesic that reaches the black hole singularity would not fall into the NS. 

Our results can be related to a recent observation which shows that in the presence of a conformally coupled scalar field, the naked singularity in $2+1$ dimensions becomes surrounded by a ``quantum dress'' \cite{dress1,Casals:2016odj,dress2,CFMZ}, which could be a physical mechanism for realizing the cosmic censorship conjecture \cite{censorship}. In the static case, the Green functions in the NS spacetime can be constructed using the method of images, since the conical singularity is obtained by identifications in AdS$_3$. The number of images to be summed over is in correspondence  with the number of self-intersections of the null geodesics calculated here. This last statement can be understood pictorially as follows: to compute Green functions one needs the geodesic distance between two points. If null geodesics in a spacetime have $N$ self-intersections, any two infinitesimally close points can be joined by $N$ topologically distinct null geodesics and to compute the two-point function, one must sum over all of these geodesics. Hence, there is a correspondence between this number $N$ and the number of images of a point under the identification used to get a NS from AdS$_3$. 

Another application lies in the study of entanglement entropy of the $1+1$ dimensional CFT dual of the NS in $2+1$ dimensions. In this analysis, according to the RT conjecture, the length of the minimal spatial geodesics play a central role \cite{RT}.  In Section 7 we computed these lengths comparing them to the corresponding geodesics in AdS$_3$, and found different results depending on which ensemble we considered: if the corresponding geodesics have the same value of $L$ the regularized length vanishes; if the corresponding geodesics have the same value of $\alpha$ the result is nonzero. Neither of these results match the finding in \cite{lengthsRT}.

There are related questions that we have not touched upon, but that could be considered in the future. For instance, a similarly detailed study of multiple conical singularities could be carried out. Moreover, if one considers identifications of AdS$_3$ different from the one considered here using a spacelike Killing vector, it is possible to build other spacetimes \cite{BHcreate,dedeo} and study their corresponding geodesics. 

\section*{Acknowledgements}
We are grateful to M. Brice\~no, M. Casals, B. Czech, A. Fabbri, G. Giribet and A. Ireland for enlightening discussions. This work has been partially funded by Fondecyt grants  1161311 and 1180368. The Centro de Estudios Cient\'{\i}ficos (CECs) is funded by the Chilean Government through the Centers of Excellence Base Financing Program of Conicyt.

\end{document}